%Paper: hep-th/9302044
%From: Corinne A. Manogue <corinne@PHYSICS.ORST.EDU>
%Date: Thu, 11 Feb 93 16:17:44 PST

%**start of header

\magnification=\magstep1       	% Imitate 12 pt output from Plain TeX.
\font\bigbold=cmbx10 scaled 1200

\newcount\EQNO      \EQNO=0
\newcount\FIGNO     \FIGNO=0
\newcount\REFNO     \REFNO=0
\newcount\SECNO     \SECNO=0
\newcount\SUBSECNO  \SUBSECNO=0
\newcount\FOOTNO    \FOOTNO=0
\newbox\FIGBOX      \setbox\FIGBOX=\vbox{}
\newbox\REFBOX      \setbox\REFBOX=\vbox{}
\newbox\RefBoxOne   \setbox\RefBoxOne=\vbox{}

%% "\normal" should be defined to restore the correct size to references which
%%           occur e.g. inside small-sized footnotes.
%%           The following line insures that a default exists.
%%           (Based on Exercise 7.7 in the TeXbook.)
\expandafter\ifx\csname normal\endcsname\relax\def\normal{\null}\fi

\def\Eqno{\global\advance\EQNO by 1 \eqno(\the\EQNO)%
    \gdef\label##1{\xdef##1{\nobreak(\the\EQNO)}}}
\def\Fig#1{\global\advance\FIGNO by 1 Figure~\the\FIGNO%
    \global\setbox\FIGBOX=\vbox{\unvcopy\FIGBOX
      \narrower\smallskip\item{\bf Figure \the\FIGNO~~}#1}}
\def\Ref#1{\global\advance\REFNO by 1 \nobreak[\the\REFNO]%
    \global\setbox\REFBOX=\vbox{\unvcopy\REFBOX\normal
      \smallskip\item{\the\REFNO .~}#1}%
    \gdef\label##1{\xdef##1{\nobreak[\the\REFNO]}}}
\def\Section#1{\SUBSECNO=0\advance\SECNO by 1
    \bigskip\leftline{\bf \the\SECNO .\ #1}\nobreak}
\def\Subsection#1{\advance\SUBSECNO by 1
    \medskip\leftline{\bf \ifcase\SUBSECNO\or
    a\or b\or c\or d\or e\or f\or g\or h\or i\or j\or k\or l\or m\or n\fi
    )\ #1}\nobreak}
\def\Footnote#1{\global\advance\FOOTNO by 1
    \footnote{\nobreak$\>\!{}^{\the\FOOTNO}\>\!$}{#1}}
\def\SameFootnote{$\>\!{}^{\the\FOOTNO}\>\!$}

\def\References{\bigskip\centerline{\bf REFERENCES}
                \smallskip\copy\REFBOX}
\def\NewRefPage{\setbox\RefBoxOne=\vbox{\unvcopy\REFBOX}
		\setbox\REFBOX=\vbox{}
		\def\References{\bigskip\centerline{\bf REFERENCES}
                		\nobreak\smallskip\nobreak\copy\RefBoxOne
				\vfill\eject
				\smallskip\copy\REFBOX}
		\def\NewRefPage{}}

%%%%%%%%%%%%%%%%%%%%%%%%%%%%%%%%%%%%%%%%%%%%%%%%%
%% end Tevian's macros for automatic numbering %%
%%%%%%%%%%%%%%%%%%%%%%%%%%%%%%%%%%%%%%%%%%%%%%%%%

\def\Order{{\cal O}}
\def\O{O}
\def\H{H}
\def\K{K}
\def\R{R}
\def\C{C}
\def\Re{{\rm Re}\;}
\def\Im{{\rm Im}\;}
\def\rhat{\hat r}
\def\shat{\hat s}
\def\that{\hat t}
\def\ihat{\vec \imath}
\def\jhat{\vec \jmath}
\def\khat{\vec k}
\def\Chi{{\bf X}}
\def\cconj{{\rm Bar}}
\def\phmsp{\mskip-14.166mu}
\let\isom = \approx

\def\2by2{$2\times 2$}
\def\det{{\rm det}}
\def\dag{^{\dagger}}

\def\noi{\noindent}

% American date ordering convention MD,Y:
\def\Atoday{\ifcase\month\or
  January\or February\or March\or April\or May\or June\or
  July\or August\or September\or October\or November\or December\fi
  \space\number\day, \number\year}
% European date ordering convention DMY:
\def\Etoday{\number\day\space\ifcase\month\or
  January\or February\or March\or April\or May\or June\or
  July\or August\or September\or October\or November\or December\fi
  \space\number\year}

% kluge to reduce size of last matrix
\def\Small{\textfont0=\sevenrm\textfont1=\seveni\textfont2=\sevensy
           \scriptfont1=\fivei}

%**end of header

%\line{\hfill \Atoday}
\line{\hfill February 9, 1993}
\line{\hfill hep-th/9302044}

\bigskip\bigskip\bigskip

\centerline{\bigbold Finite Lorentz Transformations,}
\medskip
\centerline{\bigbold Automorphisms, and Division Algebras}

\bigskip\bigskip

\centerline{Corinne A. Manogue
\Footnote{Permanent address is Oregon State University.}
}
\centerline{\it Department of Physics, Oregon State University,}
\centerline{\it	Corvallis, OR  97331, USA}
\medskip
\centerline{\it Mathematical Sciences Research Institute,
		1000 Centennial Drive,}
\centerline{\it Berkeley, CA  94720, USA}
\medskip
\centerline{\tt corinne{\rm @}physics.orst.edu}
\bigskip\bigskip
\centerline{J\"org Schray}
\medskip
\centerline{\it Department of Physics, Oregon State University,}
\centerline{\it	Corvallis, OR  97331, USA}
\medskip
\centerline{\tt schrayj{\rm @}physics.orst.edu}

\bigskip\bigskip\bigskip\bigskip
\centerline{\bf ABSTRACT}
\midinsert
\narrower\narrower\noindent
We give an explicit algebraic description of finite Lorentz transformations of
vectors in 10-dimensional Minkowski space by means of a parameterization in
terms of the octonions.  The possible utility of these results for superstring
theory is mentioned.  Along the way we describe automorphisms of the two
highest dimensional normed division algebras, namely the quaternions and the
octonions, in terms of conjugation maps.  We use similar techniques to define
$SO(3)$ and $SO(7)$ via conjugation, $SO(4)$ via symmetric multiplication, and
$SO(8)$ via both symmetric multiplication and one-sided multiplication.  The
non-commutativity and non-associativity of these division algebras plays a
crucial role in our constructions.
\bigskip
\noi

\endinsert

\vfill
\eject
%\baselineskip=2\normalbaselineskip
\baselineskip=1.3\normalbaselineskip

\Section{Introduction}

Recent research by several groups
\Ref{Some representative papers are:
\hfill\break
D.B.~Fairlie and C.A.~Manogue, {\it A parameterization of the covariant
superstring}, Phys. Rev. {\bf D36} (1987) 475;\hfill\break C.A.~Manogue and
A.~Sudbery, {\it General solutions of covariant superstring equations of
motion}, Phys.~Rev.~{\bf 40} (1989) 4073;\hfill\break F. G\"ursey, {\it Super
Poincar\'e Groups and Division Algebras}, Modern Physics {\bf A2}, (1987)
967;\hfill\break F. G\"ursey, {\it Supergroups in Critical Dimensions and
Division Algebras}, Monographs on Fundamental Physics, {\bf Proceedings of
Capri Symposia 1983-1987}, ed.\ Buccella-Franco, Lecture Notes Series No.\ 15,
American Institute of Physics, 1990, p. 529;\hfill\break I. Bengtsson \& M.
Cederwall, {\it Particles, twistors and the division algebras}, Nucl.~Phys. B
{\bf 302}, (1988) 81;\hfill\break M. Cederwall, {\it Octonionic particles and
the $S^7$ symmetry}, J. Math.  Phys. {\bf 33} (1992) 388;\hfill\break R. Foot
\& G.C. Joshi, {\it Space-time symmetries of superstring and Jordan algebras},
Int. J. Theor. Phys. {\bf 28} (1989) 1449;\hfill\break E. Corrigan and T.J.
Hollowood, {\it The Exceptional Jordan Algebra and the Superstring}, Commun.
Math. Phys. {\bf 122} (1989) 393.}\NewRefPage\hskip -.7em
 on the $(9,1)$ dimensional
\Footnote{For notational convenience we use the symbol $(m,l)$ to
denote the total dimension of Minkowski space, where $m$ is the number of
spatial dimensions and $l$ is the number of timelike dimensions.} superstring
has shown that a parameterization in terms of octonions is natural and may
help to illuminate the symmetries of the theory.  In particular, an
isomorphism between $SO(9,1)$ and $SL(2,\O)$ can be used to write the $(9,1)$
vector made up of the bosonic coordinates of the superstring as a
\2by2 dimensional hermitian matrix with octonionic entries in the same way
that the standard isomorphism between $SO(3,1)$ and $SL(2,\C)$ is used to
write a $(3,1)$ vector as a \2by2 dimensional hermitian matrix with complex
entries.  But what exactly is meant by $SL(2,\O)$?  The infinitesimal version
of $SL(2,\O)$ has been known for some time
\Ref{A. Sudbery, {\it Division algebras, (pseudo) orthogonal groups and
spinors}, J.~Phys.~A: Math.~Gen.~{\bf 17} (1987) 939.\hfill\break K.W. Chung
\& A. Sudbery, {\it Octonions and the Lorentz and conformal groups of
ten-dimensional space-time}, Phys.~Lett.~B {\bf 198}, (1987) 161.}.  However,
since the octonions are not associative, it is not possible to ``integrate''
the infinitesimal transformations to obtain a finite transformation in the
usual way.  In this paper, we show how to get around this problem and give an
explicit algebraic description of finite transformations in $SL(2,\O)$.  Along
the way, we also develop explicit octonionic characterizations of the finite
transformations of a number of other interesting groups, especially $G_2$,
$SO(7)$, and $SO(8)$.

In Section 2 we present some basic information about division algebras and
introduce our notation.  This section may be safely omitted by the reader who
is already familiar with division algebras.  In Section 3 we give an explicit
algebraic description of finite elements of $SO(3)$ and $SO(7)$.  ($SO(3)\isom
Aut(\H)$ is the group of continuous proper automorphisms of the quaternions.)
We also find a simple restriction of $SO(7)$ which gives a construction of the
continuous proper automorphisms of the octonions $G_2\isom Aut(\O)$.  Then in
Section 4 we find a related algebraic description of $SO(4)$ and two
descriptions of $SO(8)$.  We use these results in Section 5 to construct {\bf
finite} Lorentz transformations of vectors in $(5,1)$ and $(9,1)$ dimensions.
Section 6 summarizes our conclusions and discusses how our work relates to the
work of others.

\Section{Division Algebra Basics}

In this section we introduce the basic definitions and properties of the
normed division algebras.  We take an intuitive approach in order to make a
first encounter accessible.  For a more rigorous mathematical treatment see,
for example,
\Ref{R.D. Schafer, {\bf An Introduction to Non-Associative Algebras}
(Academic Press, New York, 1966).}.

According to a theorem by Hurwitz \Ref{A. Hurwitz, {\it \"Uber die Composition
der quadratischen Formen von beliebig vielen Variablen}, Nachrichten von der
Gesellschaft der Wissenschaften zu G\"ottingen, (1898) 309-316.
\hfill\break
L.E. Dickson, {\it On Quaternions and the History of the Eight Square Theorem},
Annals of Mathematics~2 {\bf 20}, (1919) 155.}, there are only four algebras
over the reals, called normed division algebras, with the property that their
norm is compatible with multiplication.  These are the reals $\R$, the
complexes $\C$, the quaternions $\H$, and the octonions $\O$; which we denote
by $\K_n$, where $n=1,2,4,8$ is their respective dimension as vector spaces
over the reals.

First we need to define these algebras.  An element $p$ of $\K_n$ is
written\Footnote{Throughout this paper summation over repeated indices is
implied unless otherwise noted.} $p=p^i e_i$ for $p^i \in \R$, where
$i=1,\dots,n$.  The $e_i$'s can be identified with an orthonormal basis in
$\R^n$, but they also carry the information which determines the algebraic
structure of $\K_n$.  Addition on $\K_n$ is just addition of vectors in
$\R^n$:

$$p+q=(p^ie_i)+(q^ie_i) = (p^i+q^i)e_i\Eqno $$\label\add

\noi
and is therefore both commutative and associative.  Multiplication is
described by the tensor $\Lambda$. ($\Lambda$ must be defined so as to contain
the structural information necessary to yield norm compatibility.  We discuss
the detailed properties of $\Lambda$ below.)

$$pq=(p^je_j)(q^ke_k) = ({\Lambda^i}_{jk}p^jq^k)e_i\Eqno$$\label\mult

\noi
where ${\Lambda^i}_{jk}\in\R {\rm\ for\ } i,j,k = 1,\ldots,n$.  We see that
multiplication is bilinear and distributive, i.e.\ determined by the products
of the basis vectors, but it is not necessarily commutative nor even
associative.

We write the multiplicative identity in $\K_n$ as $e_1=1$ and call it the real
unit.\Footnote{In most references the identity is denoted by $e_0$ or $i_0$
and indices run from $0$ through $n-1$.  For later notational convenience
our indices run from $1$ through $n$.} Due to the linearity of
\mult, $\R e_1$ is an embedding of $\R$ in $\K_n$ and multiplication with an
element of $\R\isom\R e_1$ is commutative.  The other basis vectors
satisfy $e_ie_i= e_i^2= -1= -e_1$ for $i= 2,\dots,n$ and we call them
imaginary basis units.  The imaginary basis units anticommute with each other,
i.e.\ $e_ie_j= -e_je_i$ for $i \ne j$ and the product of two imaginary basis
units yields another, i.e.\ $e_ie_j=\pm e_k$ for some $k$.

In the familiar way, we have $\{e_1= 1\}$ for $\R$ and $\{e_1= 1,e_2= i\}$
for $\C$.  For $\H$ we have $\{e_1= 1,e_2,e_3,e_4=e_2e_3\}$.  Because there
is more than one imaginary basis unit, multiplication on $\H$ is not
commutative, but it is still associative.  The rest of the multiplication
table follows from associativity.  We can visualize multiplication in $\H$ by
an oriented circle\Footnote{In the figures and occasionally in the text, we
will drop the $e$ from the notation for a basis unit and refer to it just by
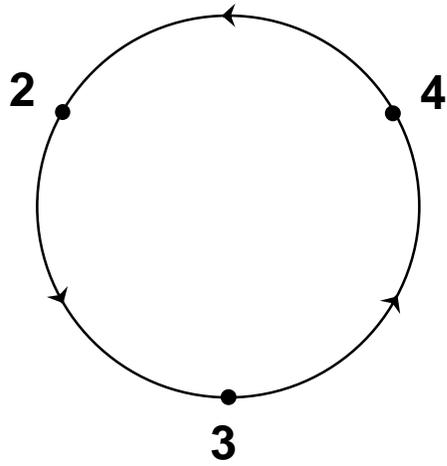
its number, i.e.\ $e_2\equiv 2$ and $e_i\equiv i$.}; see \Fig{A schematic
representation of our choice for the quaternionic multiplication table.}.
%
%\vbox to 40pt{\vfil \line{\hfil circle\hfil}\vfil}
%
%\noi
The product of two imaginary basis units, represented by nodes on the
circle, is the imaginary basis unit represented by the third node on the
line connecting them if the product is taken in the order given by the
orientation of the circle, otherwise there is a minus sign in the result.
Multiplication of the imaginary basis units in $\H$ is reminiscent of the
vector product in $\R^3:\, \ihat \times \jhat = \khat = -\jhat \times
\ihat$.  Because of this, $e_2,e_3,e_4$ are often denoted $i,j,k$.

For $\O$ the multiplication table is most transparent when written as a
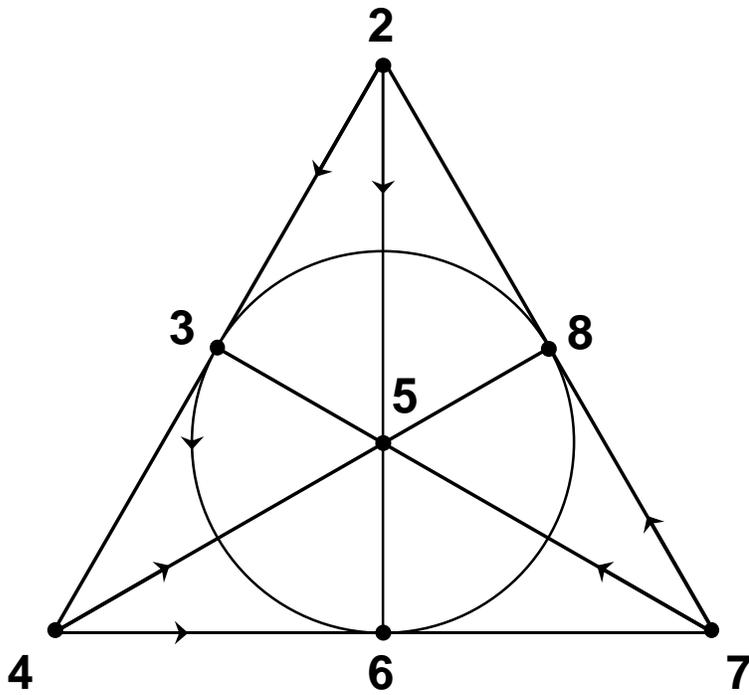
triangle; see \Fig{A schematic representation of our choice for the octonionic
multiplication table.}.
%
%\vbox to 70pt{\vfil \line{\hfil triangle\hfil}\vfil}
%
%\noi
The product of two imaginary basis units is determined as
before by following the oriented line connecting the corresponding nodes,
where each line on the triangle is to be interpreted as a circle by connecting
the ends.  Moving opposite to the orientation of the line again contributes a
minus sign, e.g.~$e_3e_4= e_2$ or $e_8e_6= -e_3$.  In general, multiplication
in $\O$ is not associative, but $e_1$ and any triple of imaginary basis units
lying on a single line span a 4-dimensional vector space isomorphic to $\H$.
Therefore products of octonions from within such a subspace are associative.
Products of triples of imaginary basis units not lying on a single line are
precisely anti-associative so switching parentheses results in a change of
sign.  For example, $e_2(e_3e_4)= e_2(e_2)= -1= (e_4)e_4= (e_2e_3)e_4$, but
$e_2(e_3e_5) = e_2(-e_7)= -e_8= -(e_4)e_5 = -(e_2e_3)e_5$.

To describe the results of switching parentheses, it is useful to define the
associator $[p,q,r]:= p(qr)- (pq)r$ of three octonions $p,q,r$.  The
associator is totally antisymmetric in its arguments.  From the antisymmetry
of the associator we see that the octonions have a weak form of associativity,
called alternativity, i.e.\ if the imaginary parts of any two of $p,q,r$ point
in the same direction in $\R^7$, the associator is zero.  In particular,
$[p,q,p]=0$.  As a consequence of alternativity, some products involving four
factors have special associativity properties given by the Moufang
\Ref{R. Moufang, {\it Zur Struktur von Alternativk\"orpern},
\hfil Mathematische Annalen, {\bf 110},
\break
(1934) 416-430.}
identities:

$$\eqalign{
q\left(p\left(qx\right)\right)&=\left(qpq\right)x\cr\noalign{\medskip}
\left(\left(xq\right)p\right)q&=x\left(qpq\right)\cr\noalign{\medskip}
q\left(xy\right)q&=\left(qx\right)\left(yq\right)\cr}
\qquad\qquad \forall p,q,x,y \in \K_n\Eqno$$\label\Moufang

As in the familiar case of the complex numbers, complex conjugation is
accomplished by changing the sign of the components of the imaginary basis
units, i.e.\ the complex conjugate of $p:=p^i e_i$ is given by

$$\overline{p}=\cconj(p):= p^1 e_1 - \sum_{i=2}^n p^i e_i\Eqno$$

\noi
We define the real and imaginary parts\Footnote{Note that $\Im p$ as we define
it is not real.  For $\H$ and $\O$ which have more than one imaginary
direction, this definition is more convenient than the usual one.} of $p$ via

$$\Re p := {1\over 2}(p + \overline{p}) \qquad{\rm and}\qquad \Im p := {1\over
2}(p - \overline{p})\Eqno$$

\noi
The complex conjugate of a product is the product of the complex conjugates in
the opposite order:

$$\overline{p q}=\overline{q}\> \overline p, \qquad\forall\,
p,q \in \K_n\Eqno$$
\label\conjswap

The inner product on $\K_n$ is just the Euclidian one inherited from $\R^n$:

$$\langle p,q\rangle= \sqrt{\sum_{i= 1}^n p^i q^i}\Eqno$$\label\inner

\noi which can be written in terms of complex conjugation via

$$\langle p,q\rangle = {1 \over 2}(p\,\overline{q}+q\,\overline{p})
={1 \over 2}(\overline{q}\,p+\overline{p}\,q)
={\rm Re} (p\,\overline{q})\Eqno$$\label\conjinner

\noi
In this language, an imaginary unit is
any vector which is orthogonal to the real unit and has norm $1$. Two
imaginary units which anticommute are orthogonal.  This geometric picture
relating orthogonality to anticommutativity is
often helpful, but it lacks the notion of associativity.

The inner product, \inner\ and \conjinner, induces a norm on
$\K_n$ given by

$$|p|=|p^ie_i| = \sqrt{\sum_{i= 1}^n (p^i)^2}=\sqrt{ p\,\overline{p}}
\Eqno$$\label\norm

\noi
It can be shown that the norm is compatible with multiplication
in $\K_n$:

$$|pq| = |p| |q|\Eqno$$\label\eightsq

\noi
In the case of the octonions, \eightsq\ is known as the eight squares theorem,
because a product of two sums, each of which consists of eight squares, is
written as a sum of eight squares.  Norm compatibility \eightsq\ and the
relation of the norm to complex conjugation \norm\ are essential for a
normed division algebra, since they allow division.  For $p \ne 0$, the
inverse of $p$ is given by

$$p^{-1}={\overline{p} \over |p|^2}\Eqno$$\label\division

An element $p\in \K_n$ can be written in exponential form just as in the
complex case:

$$p=N\exp(\theta\,\rhat)=N\left(\cos\theta +
\sin\theta\;\rhat \right)\Eqno$$\label\polar

\noi
where $N=|p| \in \R$, $\theta \in [0,2\pi)$ is given implicitly by $\Re p
=N\cos\theta$, and $\rhat$ is an imaginary unit\Footnote{We will use hats
(e.g.\ $\rhat$) to denote purely imaginary units.} given implicitly by $\Im p
=N\sin\theta\;\rhat$.  For the special case $N=1$ we will sometimes denote $p$
by the ordered pair

$$p=(\rhat,
\theta)\Eqno$$\label\polarb

What are the $m$th roots of $p=N\exp(\theta\ \rhat)\in \K_n$?  If $p$ is not
a real number, then in the plane determined by $e_1$ and $\rhat$ the
calculation reduces to the complex case, i.e.\ there are precisely
$m$ $m$th roots given by

$$p^{1\over m}=N^{1\over m} \exp\left({\theta + 2\pi l\over m} \rhat\right)
\Eqno$$\label\root

\noi
where $m\ge 2$ is a positive integer, $l<m$ is a non-negative integer, and
$N^{1\over m}$ is the positive, real $m$th root of the positive, real number
$N$.  However for $K_4$ and $K_8$, if $p\in \K_n$ is a real number the
situation is different.  If $p$ is real, it does not determine a unique
direction $\rhat$ in the pure imaginary space of $\K_n$.  Therefore
\root\ is no longer well-defined (unless, of course, the root is real).
Indeed, if ${p_{\pm}}^{1\over m}=N^{1\over m}\exp(\pm{\theta +
2\pi l\over m} e_2)$, for fixed $l$, are a complex conjugate pair of roots of
$p$ lying in $\C$, then $N^{1\over m}\exp({\theta +2 \pi l\over m} \rhat)$ is
also a root for any $\rhat$.  We see that the roots of $p$, which form complex
conjugate pairs in $\C$, in $\K_n$ form an $S^{n-2}$ subspace of
$\R^{n}$.  Throughout this paper, whenever we refer to the root of an
element of $\K_n$, we will mean any of these roots, so long as all of the
roots of that element in a given equation are taken to be the same.

In the discussion so far we assumed that the basis ${e_1,\dots,e_n}$ was
given.  But what happens if we change basis in $\K_n$?  Any linear
transformation would preserve the vector space structure of $\K_n$, but the
structure tensor $\Lambda$ would transform according to the tensor
transformation rules.  In order to preserve the multiplicative structure,
i.e.\ to get the same multiplication rules and the same formulas for complex
conjugation and norm, we would need for the transformation to be an
automorphism of $\K_n$.  Any such transformation yields a basis of the
following form: (a) $e_1$ is the multiplicative identity in $\K_n$ and must be
fixed by the transformation.  For $\R$, $\{e_1\}$ is the basis.  (b) $e_2$
can be any imaginary unit, i.e.\ anything in $\K_n$ which squares to $-1$.
For $\C$ there is only one choice (up to sign), so the basis in this case is
now complete.  (c) $e_3$ can be any imaginary unit which anticommutes with
$e_2$.  Then $e_4$, the third unit in the associative triple, is determined by
the multiplication table, i.e.\ $e_4=e_2 e_3$.  Now we have a basis for $\H$.
(d) For $\O$ we still need to pick another imaginary unit, $e_5$, which
anticommutes with all of $e_2$, $e_3$, and $e_4$.  The remaining units are
then determined by the triangle.

The procedure above provides a convenient simplification for calculations
which involve up to three arbitrary octonions $x,y,z$.  Without loss of
generality, we may assume that $x=x^1e_1 + x^2e_2$, $y=y^1e_1 + y^2e_2 +
y^3e_3$, and $z=z^1e_1 + z^2e_2 + z^3e_3 + z^4e_4 + z^5e_5$.  In particular,
any
calculation involving only one arbitrary octonion reduces to the complex case
and any involving only two arbitrary octonions reduces to the quaternionic
case.  In a calculation involving three arbitrary octonions, it may be assumed
that only one component of one of them lies outside a single associative
triple.  Only the fourth arbitrary octonion in a calculation cannot be chosen
to have some vanishing components.  These simplifications can be especially
useful when combined with computer algebra techniques.

The multiplication rules which we have chosen are not unique, but all other
choices amount to renumberings of the circle or triangle, including those
which switch signs (nodes may be relabeled $\pm 2,\ldots,\pm 8$).  Even
some of these turn out to be equivalent to the original triangle.  The
seven points of the triangle can be identified with the projective plane over
the field with two elements, so the possible renumberings of the
imaginary basis units correspond to transformations of this plane.  For future
reference we give the form of $\Lambda$ corresponding to our choice of
multiplication rules in Appendix A.

\Section{SO(n-1) and Automorphisms}

A proper automorphism $\phi$ of $\K_n$ satisfies

\def\Test{\global\advance\EQNO by 1 (\the\EQNO)%
    \gdef\label##1{\xdef##1{\nobreak(\the\EQNO)}}}

$$\eqalignno{
\phi(x+y)&= \phi(x) + \phi(y)&\Test\cr\label\linear
\phi(xy)&= \phi(x) \phi(y)\qquad {\rm (proper)}&\Test\cr%\label\proper
\noalign{\label\proper
\noi
$\forall\, x,y \in \K_n$, whereas for an improper or anti-automorphism the
order of the factors in \proper\
is reversed:}
\phi(xy)&= \phi(y) \phi(x)\qquad {\rm (improper)}&\Test\cr\label\improper}$$

\noi
{}From \conjswap\ and the non-commutativity of quaternionic and octonionic
multiplication, we see that complex conjugation is an example of an improper
automorphism for $n=4,8$.

Throughout the rest of this paper we will restrict ourselves to the set of
continuous proper automorphisms, $Aut(\K_n)$.\Footnote{All of the continuous
automorphisms of $\H$ or $\O$, including the improper ones which change the
order of the multiplication, can be obtained by taking the direct product of
$Aut(\H)$ or $Aut(\O)$ with the group $\{\bf{1},\cconj\}$.} Then \linear,
\proper, and continuity are sufficient to show that $\phi$ is a linear
transformation on $\K_n$.  As such, $\phi$ can be expressed by the action of a
real matrix ${A^i}_j$ acting on the components $x^j$ (for ${j=1,\ldots,n}$) of
$x$ viewed as a vector in $\R^n$:

$$ \phi:\,\K_n \to \K_n {\rm\ linear}
\iff\phi(x) = {A^i}_j x^je_i\Eqno$$\label\Rlinear

Combining this form of $\phi$ with the condition \proper\ and using the
multiplication rule \mult\ we obtain the following equation for the
${A^i}_j$'s:

$${A^i}_l{\Lambda^l}_{jk}={A^l}_j{\Lambda^i}_{lm}{A^m}_k
\Eqno$$\label\liegroup

\noi
This equation defines the Lie group of automorphisms in terms of $n\times n$
matrices and the structure constants of $\K_n$.

The formulation which we have just described is the usual one for Lie groups,
but it does not take advantage of the special algebraic structure of $\K_n$.
The approach which we prefer to take in this paper is to find algebraic
operations on $\K_n$ which yield maps that satisfy \linear --\proper\ without
resorting to the matrix description.  The algebraic operations which we will
find turn out to have many interesting properties.

Motivated by the structure
of inner automorphism on division rings, let us consider conjugation maps
$\phi_q$ on $\K_n= \H, \O\; (n= 4, 8)$ for $q \in {\K_n}^* = \K_n - \{0\}:$

$$\phi_q:\vtop{\halign{$\hfil#\hfil$&$\hfil{}#{}\hfil$&$#\hfil$\cr
\K_n&\to&\K_n\cr
\noalign{\smallskip}
x&\mapsto&q x q^{-1}\cr}}\Eqno$$\label\multconj

\noi
These maps are well-defined even for $\K_8 = \O$ since the associator
$[q,x,q^{-1}]$ vanishes.  (This vanishing associator also implies
that $(\phi_q)^{-1} = \phi_{q^{-1}}$
and $(\phi_q)^2 = \phi_{q^2}$ for both $\H$ and $\O$).
The maps \multconj\ satisfy \linear\ and fix the real part of $x$.

We see from \multconj\ that a rescaling of $q$ does not effect the
transformation, so without loss of generality we may divide out the
multiplicative center, $\R^{\,*}=\R-\{0\}$, and consider only $q$'s of unit
norm, i.e.\ $q=(\rhat, \theta)$.\Footnote{We could also identify antipodal
points on the unit sphere $(S^{n-1})$, since $\phi_q =
\phi_{-q}$.}\label\antipodal\
Notice that now $q^{-1}=\overline{q}$.
Thus we have a map $\Phi$ which takes $\{q \in \K_n : \, |q| = 1\} \isom
{\K_n}^* / \R^{\,*} \isom
S^{n-1}$ to
$\{\phi_q\}$, where $\phi_q$ is a linear transformation on $\K_n$:

$$\Phi:\vtop{
\halign{$\hfil#$&$\hfil{}#{}\hfil$&$#\hfil$\cr
\{q \in \K_n : \, |q| = 1\}&\to&L(\K_n,\K_n)\cr
\noalign{\medskip\smallskip}
q&\mapsto&\phi_q=\phi_{(\rhat,\theta)}:
\vtop{\halign{$\hfil#\hfil$&$\hfil{}#{}\hfil$&$#\hfil$\cr
\K_n&\to&\K_n\cr
\noalign{\smallskip}
x&\mapsto&q x\overline{q}=\exp(\theta\,\rhat)\, x \exp(-
\theta\,\rhat)\cr}}\cr}}\Eqno$$\label\rhatconj

\noi
We see from \eightsq\ that $\phi_q$ is an isometry:

$$|\phi_q(x)| = |q| |x| |\overline{q}| = |x|\Eqno$$\label\isometry

\noi
In particular it leaves the norm of the imaginary part
invariant so the associated $n\times n$ matrix $A_q$
(which is defined by: $\phi_q(x) = {(A_q)^i}_j x^j e_i$)
is orthogonal and splits into a trivial
$1 \times 1$ block for the real part and an $(n-1)\times (n-1)$ block $R_q$
which lies in $SO(n-1)$.  The determinant of $A_q$ is positive, because
$
\phi_q =(\phi_{\sqrt{q}})^2$ (equivalently
$A_q = (A_{\sqrt{q}})^2$).

Now we
will study the structure of $\Phi(S^{n-1})$ by looking at generic
examples of maps $\phi_q$.

\Subsection{\it Quaternions and $SO(3)$:}

For $\K_1=\R$ and $\K_2=\C$, multiplication is commutative and the conjugation
maps \multconj\ are trivial.  Therefore let us examine the first nontrivial
case, $\K_4=\H$.  If we consider, for example, $\rhat = e_2$, we get

$$\exp(\theta\, e_2)\, x \exp(-\theta\, e_2) = x^1 e_1 + x^2 e_2 +
(\cos 2\theta\; x^3 -\sin 2\theta\; x^4) e_3 + (\sin 2\theta\; x^3 +
\cos 2\theta\; x^4) e_4$$
$$\iff A_{(e_2,\theta)} = \left [\matrix{1&0&0&0\cr 0&1&0&0\cr
0&0&\cos 2\theta&-\sin 2\theta\cr 0&0&\sin 2\theta&\phantom{-}
\cos 2\theta\cr}\right ]\quad
{\rm so}\quad
R_{(e_2,\theta)} = \left [\matrix{1&0&0\cr
0&\cos 2\theta&-\sin 2\theta\cr
0&\sin 2\theta&\phantom{-}\cos 2\theta\cr}\right ]\Eqno$$

\noi
This is just a rotation of the imaginary part of $x$ around $e_2$ by an angle
of $2\theta$, i.e.\ it is a rotation in the $3$-$4$ plane.
Similarly, we see that $\phi_q$ with $q = \exp(\theta\,\rhat)$,
for any imaginary unit $\rhat$,
is a rotation of the imaginary part of $x$ around $\rhat$ by an angle of $2
\theta$ .
Thus $\Phi$ is the universal covering map, mapping $S^3$ onto $SO(3) \isom
Aut(\H)$.  Since multiplication in $\H$ is associative, composition of maps is
given by multiplication in $\H$, i.e.\ $\phi_p \circ
\phi_q = \phi_{pq}$ (equivalently $A_p A_q = A_{pq}$),$\; \forall\, p,q \in \H
{\rm\ with\ } |p| = |q| = 1$.  Therefore, $\Phi$ is also a group
homomorphism.\Footnote{One application of this homomorphism is a quick
derivation of the expression for the composition of two rotations given in
terms of axes and angles of rotation. If $p = \exp(\theta\,\rhat)$ and $q =
\exp(\eta\,\shat)$, then $pq =
\exp(\zeta\,\that)$ where $\that = {\Im(pq)/ |\Im(pq)|}$ and $\cos\zeta =
\Re(pq)$.  So a $2
\eta$ rotation around $\shat$ followed by a $2 \theta$ rotation around $\rhat$
is the same as a $2 \zeta$ rotation around $\that$.}

We have just parameterized rotations in the 3-dimensional purely imaginary
subspace of the quaternions by fixing an axis of rotation and then specifying
the value of a continuous parameter, the angle $\theta$, which describes the
amount of the rotation around that axis in the unique plane orthogonal to that
axis.  We call this parameterization the axis-angle form. But in dimension
greater than 3, there is no unique plane orthogonal to a given axis.
Therefore in the octonionic case it will not be sufficient to specify a
rotation axis and an angle of rotation.  Instead, we will parameterize
rotations in another way, which we first describe here for the quaternionic
case.

To accomplish a given elementary rotation (a rotation which takes place
in a single coordinate plane), we use a composition of two particular
axis-angle rotations, which we call flips because they are both rotations by
the same constant angle $\pi$.  The angle $\theta$ between the axes of the
two flips then takes on the role of a continuously changing parameter
which describes the magnitude of the combined rotation.  Specifically, choose
any two anticommuting (i.e.\ perpendicular) imaginary units $\rhat$ and $\shat$
which lie in the plane of the desired rotation.  Then if the desired amount of
rotation in that plane is $2\theta$, do two flips around the two directions
$\rhat$ and $\cos\theta\;\rhat +
\sin\theta\;\shat$  (which are separated by the angle $\theta$).  To do this,
we define the composition $\phi_{(\rhat,\shat,\theta\vert\alpha)}^{(2)}$ via

$$\phi_{(\rhat,\shat,\theta\vert\alpha)}^{(2)}
:= \phi_{(\cos\theta\;\rhat + \sin\theta\;\shat,\alpha)}
\circ \phi_{(\rhat,-\alpha)}\Eqno$$\label\phitwoalpha

\noi
in particular, for $\alpha = {\pi\over 2}$:

$$\eqalign{\phi&_{(\rhat,\shat,\theta\vert {\pi\over 2})}^{(2)}(x):=\cr
&\exp \left({\pi\over 2}(\cos\theta\;\rhat + \sin\theta\;\shat)\right)
\left[\exp\left(-{\pi \over 2}\,\rhat\right) x
\exp\left({\pi \over 2}\,\rhat\right) \right]
\exp\left(-{\pi\over 2}(\cos\theta\;\rhat + \sin\theta\;\shat)\right)
\cr}\Eqno$$\label\phitwo

\noi
where the superscript ``(2)'' indicates the number of simple axis-angle
$\phi$'s involved in the composition.
In order to
understand why $\phi^{(2)}$ works, consider its effects on different
subspaces.  In the plane spanned by $\rhat$ and $\shat$, $\phi^{(2)}$ is just
the composition of two reflections with respect to the two directions $\rhat$
and $\cos\theta\;\rhat + \sin\theta\;\shat$ as mirror lines, amounting to a
total rotation by $2\theta$, so that $\theta$ is indeed the continuously
changing parameter.  In particular $\phi_{(\rhat,\shat,0)}^{(2)} = \bf{1}$.
In the direction orthogonal to the plane, the flips are in opposite directions
and therefore cancel.  We call $\phi^{(2)}$ the plane-angle form of the
rotations because it parameterizes rotations in terms of their plane and
angle.  In the case of the quaternions we can of course use the group
homomorphism property of the $\phi$'s to express $\phi^{(2)}$ as a single
$\phi$:

$$\phi_{(\rhat,\shat,\theta | {\pi \over 2})}^{(2)}
= \phi_{(\cos\theta\;\rhat + \sin\theta\;\shat,{\pi\over 2})}
\circ \phi_{(\rhat,-{\pi \over 2})}
= \phi_{(\rhat\shat,\theta)}
\Eqno$$
\noi since
$$\exp\left({\pi\over 2}(\cos\theta\;\rhat + \sin\theta\;\shat)\right)
\exp\left(-{\pi \over 2}\rhat\right)
=(\cos\theta\;\rhat + \sin\theta\;\shat)(-\rhat)
=\cos\theta + \sin\theta\;\rhat\shat\Eqno$$

\noindent
We see that $\phi_{(\rhat,\shat,\theta| {\pi \over 2})}^{(2)}$ only depends
on the product $\rhat\shat$, which in turn depends only on the plane (and
orientation) of $\rhat$ and $\shat$.  Therefore any pair of anticommuting
units spanning the same plane with the same orientation may replace $\rhat$
and $\shat$ without changing the combined transformation.

We have seen that $\Phi$ maps all of $S^3$ to $Aut(\H)$, but this new
parameterization of the rotations only uses $q$'s of the form
$\exp\left({\pi\over 2}\rhat\right)$, i.e.\ the angle in each of the
individual flips is always the constant ${\pi \over 2}$.\Footnote{Because
$(-\theta)\rhat$ can be interpreted as $\theta (-\rhat)$, the choice of the
sign of the angle in each flip has no consequences.  Therefore we have chosen
the signs in \phitwo\ (and in later sections) for convenience.} This means that
just
a single $S^2$ slice of $S^3$ (the equator) maps under $\Phi$ to a
generating set for $Aut(\H)$.

\Subsection{\it Octonions and $SO(7)$:}

Now let us examine the more complicated case, $\K_8=\O$.  We notice that for
the octonions each line in the triangle, and more generally each associative
triple of anticommuting, purely imaginary octonions of modulus 1, is just a
copy of the imaginary units $\{e_2,e_3,e_4\}$ in $\H$.  Therefore, if we
consider the same conjugation map as we did in the quaternionic case with
$q=\exp(\theta\, e_2)$, we obtain the associated matrix $A_{(e_2,\theta)}$:

$$
A_{(e_2,\theta)}
=\left
[\matrix{1&0&0&0&0&0&0&0\cr 0&1&0&0&0&0&0&0\cr 0&0&\cos 2\theta&-\sin
2\theta&0&0&0&0\cr 0&0&\sin 2\theta&\phantom{-}\cos 2\theta&0&0&0&0\cr
0&0&0&0&\cos 2\theta&-\sin 2\theta&0&0\cr
0&0&0&0&\sin 2\theta&\phantom{-}\cos 2\theta&0&0\cr
0&0&0&0&0&0&\cos 2\theta&-\sin 2\theta\cr
0&0&0&0&0&0&\sin 2\theta&\phantom{-}\cos 2\theta\cr}\right ]\Eqno$$

\noi
We see that this transformation yields three simultaneous rotations by an
angle of $2 \theta$ in three mutually orthogonal planes which are all
orthogonal to $e_2$.  The pairs of imaginary units which are rotated into each
other are just the pairs which each form an associative triple with $e_2$.
Moreover, since the rotations in the three planes are equal, the choice of
these planes is not unique.

For an arbitrary $\rhat$ we can always find a (nonunique) set of 3 pairwise
orthogonal planes, orthogonal to $\rhat$, such that $\phi_{(\rhat, \theta)}$
represents an axis-angle rotation in each of the quaternionic subspaces
spanned by one of the planes and $\rhat$.  For the special case $\theta = {\pi
\over 2}$, $A_{(\rhat, \theta)}$ has 8 real eigenvalues, 6 of which are $-1$.
In this case the extra degeneracy means that if we choose $\rhat$ anywhere on,
for example, the 2-3-4 subspace the effect on the 5-6 and 7-8 planes is the
same.

Because each $\phi_q$ rotates three planes, it looks naively as if we should
only be able to describe a subset of $SO(7)$ in this way.  Surprisingly, this
is not true.  We can in fact describe all of $SO(7)$ and it turns out that the
non-associativity of multiplication in $\O$ plays a crucial role.  For $\K_8
=\O$, $\phi_p \circ \phi_q \ne \phi_{pq}$ in general, i.e.\ $\Phi$ is not a
group homomorphism.  In fact, $\phi_p \circ \phi_q \ne \phi_{r}$, for {\bf
any} $r \in \O$ unless $\Im p$ and $\Im q$ point in the same direction.  It is
this fact which allows $\Phi(S^7)$ to generate a Lie group with dimension
larger than $7$.  For instance, by using more than one mapping, we can give
explicit expressions for all of the elementary rotations.  An elementary
rotation in the $i$-$j$ plane, for example, is given by $\phi_p \circ
\phi_{\,\overline{q}} \circ \phi_p \circ \phi_q$, where $q = \exp(\theta\,
e_k)\,
,\;p=\exp\left({\pi \over 2} e_i\right)\, ,\; e_k = e_i e_j$.  This yields a
rotation by $4 \theta$ in the $i$-$j$ plane.  The extra transformations undo
the rotation in the other two planes, which were initially rotated by
$\phi_q$.  The elementary rotations generate all of $SO(7)$.

Alternatively, the plane-angle form of the quaternionic case (involving only
rotations with $\theta= {\pi\over 2}$) goes through as before, since in all
the directions orthogonal to both axes the two rotations by $\pi$ still
cancel.  Therefore $\phi_{(e_i,e_j,\theta\vert{\pi\over 2})}^{(2)}$ is another
way of expressing a rotation by $2\theta$ in the $i$-$j$ plane.  We see from
the axis-angle form of the rotations that $\Phi$ maps the unit sphere in $\O$
to a generating set of $SO(7)$.  As the plane-angle form shows, the equatorial
$S^6$ is actually sufficient to provide a generating set of $SO(7)$.

\Subsection{\it Octonions and $G_2$:}

In the octonionic case we have obtained a larger group than we were looking
for; all of $SO(7)$ instead of only its subgroup (of automorphisms of the
octonions) $G_2$.  However, we shouldn't have expected $\phi_q$ to be an
automorphism since
\proper\ is equivalent to

$$(q x q^{-1})(q y q^{-1}) = q (xy)q^{-1}\Eqno$$\label\multoct

\noi
which would require the $q$'s in between $x$ and $y$ to cancel.  \multoct\ only
holds in general if multiplication is associative; but for certain
choices for $q$, $\phi_q$ might still be an automorphism.  For $q =
\exp(\theta\, e_2)$, we find that \multoct\
places no restriction on $\theta$ if $e_2$, $\Im x$, and $\Im y$ lie on one
line in the triangle (when the calculation reduces to the quaternionic case).
However, if $e_2$, $x$, and $y$ contain anti-associative components, their
products are not equal on the two sides of \multoct.  Instead we obtain the
following two equations for $\theta$:

$$\vcenter{\halign{$\hfil#$&$#\hfil$\cr \cos 4 \theta &{}= \cos 2 \theta \cr
-\sin 4 \theta &{}= \sin 2 \theta\cr}}\Eqno$$\label\angle

\noi
The solutions for \angle\ are $\theta = k {\pi\over 3}$, $k=0,
\dots, 5$.  Obviously, $e_2$ can be replaced by any purely imaginary
octonionic unit.  Hence a single mapping, $\phi_q$, is an automorphism of $\O$
if and only if

$$q=\exp\left(k {\pi\over 3}\,\rhat\right),\qquad k=0, \dots,
5\Eqno$$\label\sixthroot

\noi
i.e.\ if and only if $q$ is a sixth root of unity, $q^6 = 1$.

These maps are not all of the automorphisms of $\O$, but they do generate the
whole group.  As in the previous section, we need to consider compositions of
$\phi_q$'s, this time satisfying \sixthroot.  We will show that we can obtain
all of $G_2$ in this way by checking that the dimension of the associated Lie
algebra is correct.  Notice that the set of allowed $q$'s splits into four
pieces depending on the value of $\Re q$, $\{\Re q =
\pm 1,\pm {1\over 2}\}$.  If $q = \pm 1$, then $\phi_q$ is the identity.
The piece with $\Re q = -{1\over 2}$ is made up of points which are antipodal
in $S^7$ to the piece with $\Re q = {1\over 2}$ (see Footnote 9).  Therefore
these two pieces contain the same maps and we
only need to consider the piece with $\Re q = {1 \over 2}$.

To determine the
group that is generated by these maps, we consider compositions of maps of the
form $\phi_{(i,j,\theta\vert{\pi\over 3})}^{(2)}$.  These are flips involving
angles of ${\pi\over 3}$ so that each individual $\phi$ is an automorphism
(instead of ${\pi\over 2}$ as in the last section).  Of course,
$\phi_{(i,j,0\vert{\pi\over 3})}^{(2)} = \bf{1}$.  Since $(\phi_q)^{-1} =
\phi_{q^{-1}}$, we also see that the set of maps with $\Re q = {1 \over
2}$ contains the inverse of each element.  A dimensional analysis
of the associated Lie algebra
finds the dimension of the space spanned by

$$\left\{ {\rm{d}
\over {\rm{d}
\theta}} \phi_{(i,j,\theta | {\pi \over 3})}^{(2)} {\Big |}_{\theta = 0}:
\;{{i,j =
2,\ldots,8}, i \ne j} \right\}\Eqno$$\label\Lie

\noi
to be 14 as follows.  There are $7 \times 6=42$ choices for $i$ and $j$.  It
turns out that the 6 choices belonging to one associative triple of units only
give 3 linearly independent generators, which leaves us with 21. In addition
three triples which have one unit in common also share one generator, which
cuts the number down by 7 leaving us with 14 independent generators for the
Lie algebra.\Footnote{To do this analysis we returned to the matrix
representation of $G_2$, \liegroup, and used the computer algebra package
MAPLE.  The calculations are nontrivial, especially the proof that the
remaining 14 generators are really independent.  We were surprised by the
result that the generator of $\phi_{(i,j,\theta | {\pi \over 3})}^{(2)}$ is
not simply related to the generator of $\phi_{(j,i,\theta | {\pi \over
3})}^{(2)}$.} Therefore the group generated is a 14-dimensional subgroup of
$G_2$, i.e. $G_2$ itself.

{}From the form of $\phi_{(i,j,\theta\vert{\pi\over 3})}^{(2)}$ we see that
$\left\{\phi_q:\; q=\exp\left({\pi \over 3}\, \rhat\right) = {1 \over 2} +
{\sqrt{3} \over 2}\,
\rhat\right\}
\allowbreak\isom S^6$ actually suffices as generating set for $G_2$.
We saw in the previous subsection that $\Phi$ maps the equatorial $S^6$ to a
generating set of $SO(7)$.  Here we see that
$\Phi$ maps a different $S^6$ slice of the octonionic unit sphere to a
generating set of $G_2$.

\Subsection{\it Some Interesting Asides:}

As an interesting aside, we derive two new identities for commutators in $\O$
in the following way.  Let $q = {1 \over 2} + {\sqrt{3} \over 2}\,\rhat$ in
\multoct.  Then the terms containing $\sqrt{3}$ and not containing it must be
equal independently.  Thus we obtain

$$\vcenter{\halign{$\hfil#$&$#\hfil$\cr 4[\rhat,xy]&{}= (x - 3 (\rhat x
\rhat))[\rhat,y] + [\rhat,x](y - 3 (\rhat y \rhat)) \cr \noalign{\smallskip}
[\rhat,x][\rhat,y]&{}= xy - 4 \rhat (xy) \rhat + x (\rhat y \rhat) + (\rhat x
\rhat)y - 3 (\rhat x \rhat) (\rhat y \rhat)\cr}} \Eqno$$\label\comm

\noi where $x,y,\rhat\in \O$ with $\Re \rhat = 0\, ,\; |\rhat| =1$.

As another interesting aside, we note that if $q^6 = 1$ then $q^3 = \pm 1$
which implies ${\phi_q}^3 = \bf{1}$.  This means that the set of elements of
$G_2$ which are third roots of the identity generate $G_2$, because it
contains all of the maps $\phi_q$ with $q^6 = 1$.  But there are third roots
of the identity map which are not given by any single $\phi_q$ with $q$ in
$\O^{*}/\R^{*}$.  This is due to the fact that $\phi_q$ is determined
completely by its fixed direction $\rhat$, whereas a third root of the
identity map has more free parameters.  For example, the following matrix is
associated with an automorphism of $\O$ which fixes $e_2$ and its third power
is the identity, but it is not equal to $A_q$ with $q =
\exp\left({\pi \over 3} (\pm e_2)\right)$:

$$\left [\matrix{1&0&0&0&0&0&0&0\cr
0&1&0&0&0&0&0&0\cr
0&0&\cos {2\pi \over 3}&-\sin {2\pi \over 3}&0&0&0&0\cr
0&0&\sin {2\pi \over 3}&\phantom{-}\cos {2\pi \over 3}&0&0&0&0\cr
0&0&0&0&1&0&0&0\cr
0&0&0&0&0&1&0&0\cr
0&0&0&0&0&0&\phantom{-}\cos {2\pi \over 3}&\sin {2\pi \over 3}\cr
0&0&0&0&0&0&-\sin {2\pi \over 3}&\cos {2\pi \over 3}\cr}\right ]\Eqno$$

A similar statement holds for the generating set of $SO(7)$ which we found.  It
contains maps which square to the identity, because we had $q = \exp\left({\pi
\over 2}\, \rhat\right)$ whence $q^2=-1$.  But again not all the elements of
$SO(7)$
which square to the identity are given as a $\phi_q$.

\Section{More Isometries}

Due to \eightsq , we see that multiplying an element of $\H$ or $\O$ by an
element of modulus 1 is always an isometry.  The isometries of the previous
section ($SO(n-1)$ and $Aut(K_n)$ for $n=4,8$) were all obtained using the
asymmetric product, $\phi_q(x)=q x q^{-1}$.  In this section we examine two
other classes of isometries on $\H$ and $\O$.

\Subsection{\it Symmetric Products:}

First we show that it is possible to describe all of $SO(n)$ for $n=4,8$ using
symmetric products.  We define

$$\Psi:
\vtop{\halign{$\hfil#$&$\hfil{}#{}\hfil$&$#\hfil$\cr
\{q \in \K_n : \, |q| = 1\}&\to&L(\K_n,\K_n)\cr
\noalign{\medskip\smallskip}
q&\mapsto&\psi_q=\psi_{(\rhat,\theta)}:
\vtop{\halign{$\hfil#\hfil$&$\hfil{}#{}\hfil$&$#\hfil$\cr
\K_n&\to&\K_n\cr
\noalign{\smallskip}
x&\mapsto&q x q=\exp(\theta\,\rhat)\,
x\, \exp(\theta\,\rhat)\cr}}\cr}}\Eqno$$\label\symmult

\noi
As with the conjugation maps, this is
well-defined even for $\K_8 = \O$, since the associator $[q,x,q]$ vanishes.
As before $(\psi_q)^{-1} =
\psi_{q^{-1}}$ and $(\psi_q)^2 = \psi_{q^2}$ hold.  We also note that $\psi_q
= \psi_{-q}$ and that $\psi_q$ is linear.

This isometry, however, does not
fix the reals.  We denote the matrix associated with $\psi_q$ by $B_q$,
where $\psi_q(x) = {{(B_q)}^i}_j x^j e_i$.  Then $B_q \in SO(n)$ since
$\psi_q = (\psi_{\sqrt{q}})^2$ (equivalently, $B_q =
(B_{\sqrt{q}})^2$).
Letting $q = \exp(\theta\, e_2)$, we obtain

$$B_{(e_2,\theta)} = \left [\matrix{\cos 2\theta&-\sin 2\theta&0&\ldots&0\cr
\sin 2\theta&\phantom{-}\cos 2\theta&0&\ldots&0\cr
0&0&1&\ldots&0\cr
\vdots&\vdots&\vdots&\ddots&\vdots\cr 0&0&0&\ldots&1\cr}\right ]\Eqno$$

\noi This is just a rotation by $2 \theta$ in the $1$-$2$ plane.  Similarly,
any rotation by $2 \theta$ in the plane spanned by $e_1$ and any imaginary unit
$\rhat$ is given by  $\psi_q$ with $q = \exp(\theta\, \rhat)$.

But what about rotations in the purely imaginary subspace, $SO(n-1)$?
Recall from the last section that the plane-angle construction of the
elementary rotations in $SO(n-1)$ used a composition of two flips $\phi_p
\circ \phi_q$ where $p$ and $q$ were both purely imaginary.  But notice that
$\psi_q = -\phi_q$ when $q$ is imaginary, i.e.\ when $\theta = {\pi \over 2}$.
Thus the maps $\{\psi_q:\; q=\exp\left({\pi \over 2}\,
\rhat\right),\,\Re\rhat=0,\, |\rhat|=1 \}$ generate a group which includes
$SO(n-1)$.  Since we already found the rotations involving the real part we
see that $\Psi(S^{n-1})$ generates all of $SO(n)$.

It is worth noting that the $\psi_q$'s work differently from the $\phi_q$'s.
For a single $\psi_q$, $q$ is in the plane of rotation, whereas for a single
$\phi_q$, $q$ was a fixed direction.  Also, $\psi_p \circ \psi_q(x) = p(q x
q)p \ne \psi_{pq} = (pq)x(pq)$, even for $\H$, since the order of the products
is different.  Therefore $\Psi$ is not a group homomorphism.

However the Moufang identities \Moufang\ do demonstrate a partial group
homomorphism property by providing a way of combining three
$\psi$'s together into a single $\psi$ in some cases.  For arbitrary $p,q \in
\K_n$, with $|q| = |p| = 1$,

$$\psi_q\circ\psi_p\circ\psi_q = \psi_{qpq}\qquad {\rm since} \qquad
q\left(p\left(qxq\right)p\right)q =\left(qpq\right) x \left(qpq\right)\;\;
\forall x \in \K_n\Eqno$$\label\psiqpq

For any anticommuting imaginary units $\rhat$ and $\shat$, the following
identity is straightforward to prove:

$$\exp(\theta\,\shat)=
\exp\left(-{\pi \over 4}\, \rhat\right) \exp\left({\pi \over 2}(\cos \theta\;
\rhat + \sin \theta\;\shat)\right) \exp\left(-{\pi \over 4}\,\rhat\right)
\Eqno$$\label\pifourth

\noi
Together with \psiqpq, \pifourth\ shows that
a rotation $\psi_{(e_i,\theta)}$ in the $1$-$i$ plane by an arbitrary angle
$2\theta$ can be described as a combination of flips of fixed angle:

$$\psi_{(e_i,\theta)}=\psi_{(\rhat,-{\pi \over 4})} \circ \psi_{(\cos
\theta\;\rhat + \sin \theta\;e_i,{\pi \over 2})} \circ \psi_{(\rhat,-{\pi
\over 4})}\Eqno$$\label\psiflips

\noi
where $\rhat$ is any imaginary unit which anticommutes with $e_i$. \psiflips\
uses flips of angle ${\pi \over 2}$ and ${\pi \over 4}$. But since a flip with
an angle of ${\pi\over 2}$ can be written as the square of a flip with angle
${\pi\over 4}$ and since we were able to write $SO(n-1)$ in terms of flips
with angle ${\pi\over 2}$, we can write all of $SO(n)$ in terms of flips of
fixed angle ${\pi\over 4}$.  Therefore the image under $\Psi$ of an $S^{n-2}
\isom \{q =\exp\left({\pi\over 4}\, \rhat\right):\,\Re\rhat=0,\,|\rhat|=1\}$
slice of $S^n$
suffices to generate all of $SO(n)$.

To understand how \psiflips\ works, notice that the first flip rotates
the real direction into some fairly arbitrary imaginary direction $\rhat$.
The second flip then rotates this imaginary direction $\rhat$ with the
physically significant imaginary direction $\shat$.  The last flip rotates the
former real part back into place\Footnote{This sounds much like manipulations
of the Rubik's Cube, which indeed inspired JS in part.}.

\Subsection{\it One-sided Multiplication:}

Now we consider one-sided multiplication.  Of course, left multiplication and
right multiplication with elements of modulus 1 together generate $SO(n)$
because, in particular, they generate the $\psi_q$'s.  But what about left
multiplication alone?  We define

$$\Chi:
\vtop{\halign{$\hfil#$&$\hfil{}#{}\hfil$&$#\hfil$\cr
\{q \in \K_n : \, |q| = 1\}&\to&L(\K_n,\K_n)\cr
\noalign{\medskip\smallskip}
q&\mapsto&\chi_q=\chi_{(\rhat,\theta)}:\vtop{\halign
{$\hfil#\hfil$&$\hfil{}#{}\hfil$&$#\hfil$\cr
\K_n&\to&\K_n\cr
\noalign{\smallskip}
x&\mapsto&q x=\exp(\theta\,\rhat)\, x\cr}}\cr}}\Eqno$$\label\leftmul

\noi
For both $\H$ and $\O$, we have $(\chi_q)^{-1} = \chi_{q^{-1}}$ and
$(\chi_q)^2 = \chi_{q^2}$, since the associators $[q^{-1},q,x]$ and $[q,q,x]$
vanish.  The following relation, connecting the maps $\phi_q$ and $\psi_q$
with $\chi_q$, holds for the same reason:

$$\chi_q = \phi_{\sqrt{q}} \circ \psi_{\sqrt{q}} = \psi_{\sqrt{q}} \circ
\phi_{\sqrt{q}}\Eqno$$\label\chiphipsi

\noi
Of course we can no longer identify antipodal points since $\chi_{-q} = -
\chi_q \ne \chi_q$.

For the quaternions $\Chi$ is a group homomorphism, $\chi_p \circ
\chi_q = \chi_{pq}$.  So $\Chi(S^3)$ must be a 3-dimensional subgroup of
$SO(4)$.  Therefore, to investigate the structure of any $\chi_q$ on $\H$, it
will be sufficient to consider $\chi_q$ with $q = \exp(\theta\, e_2)$.  The
associated matrix $C_{(e_2,\theta)}$ is

$$C_{(e_2,\theta)} = \left [\matrix{\cos \theta&-\sin \theta&0&0\cr
\sin \theta&\phantom{-}\cos \theta&0&0\cr
0&0&\cos \theta&-\sin \theta\cr
0&0&\sin \theta&\phantom{-}\cos \theta\cr}\right ]\Eqno$$

\noi
This transformation rotates two orthogonal planes by $\theta$.  For the
general case $q=\exp(\theta\,\rhat)$, the rotations are in the plane spanned
by $e_1$ and $\rhat$ and the plane orthogonal to that, as can be seen from the
relation \chiphipsi\ and our previous investigation of maps $\phi_q$ and
$\psi_q$.

It is interesting that $\Chi(S^3)$ is {\bf not} $SO(3)$, much less $SO(4)$.
We might expect, then, that left multiplication for $\K_8=\O$ would only
describe a subgroup of $SO(8)$.  Surprisingly this is not the case.  It turns
out that the non-associativity of octonionic multiplication allows left
multiplication to generate all of $SO(8)$, as follows:

First we consider $\chi_{(e_2,\theta)}$.  The associated matrix
$C_{(e_2,\theta)}$ is:

$$C_{(e_2,\theta)}
= \left [\matrix{\cos \theta&-\sin \theta&0&0&0&0&0&0\cr
\sin \theta&\phantom{-}\cos \theta&0&0&0&0&0&0\cr
0&0&\cos \theta&-\sin \theta&0&0&0&0\cr
0&0&\sin \theta&\phantom{-}\cos \theta&0&0&0&0\cr
0&0&0&0&\cos \theta&-\sin \theta&0&0\cr
0&0&0&0&\sin \theta&\phantom{-}\cos \theta&0&0\cr
0&0&0&0&0&0&\cos \theta&-\sin \theta\cr
0&0&0&0&0&0&\sin \theta&\phantom{-}\cos \theta\cr }\right ]\Eqno$$
\label\chietwo

\noi
$\chi_{(\rhat, \theta)}$ always rotates four planes by an angle $\theta$.
(This is also clear from \chiphipsi\ and the results of previous sections.)

Now suppose we want to do an elementary rotation in just one
of these four planes.  The key idea is that the composition of two maps (c.f.\
{\phitwo})

$$\chi_{(\shat,\that,\theta\vert{\pi\over 2})}^{(2)}(x):= \exp\left(
{\pi \over 2} (\cos\theta\; \shat +
\sin\theta\; \that)\right) \left[\exp\left(-{\pi \over 2} \shat\right)\; x
\right]\Eqno$$\label\chitwo

\noi
where $\shat \that=\rhat$, will rotate exactly the same four planes as the
map $\chi_{(\rhat, \theta)}$, but because of non-associativity the rotations
will not all be in the same direction in both cases.
In particular, the parts of $x$ which
anti-associate with $s$ and $t$ will be rotated in opposite directions in the
two cases.

As an example, consider
$C_{(3,4,\theta)}^{(2)}$, the matrix associated with
$\chi_{(3,4,\theta\vert{\pi\over 2})}^{(2)}$:

$$C_{(3,4,\theta)}^{(2)} = \left
[\matrix{\cos \theta&-\sin \theta&0&0&0&0&0&0\cr
\sin \theta&\phantom{-}\cos \theta&0&0&0&0&0&0\cr
0&0&\cos \theta&-\sin \theta&0&0&0&0\cr
0&0&\sin \theta&\phantom{-}\cos \theta&0&0&0&0\cr
0&0&0&0&\phantom{-}\cos \theta&\sin \theta&0&0\cr
0&0&0&0&-\sin \theta&\cos \theta&0&0\cr
0&0&0&0&0&0&\phantom{-}\cos \theta&\sin \theta\cr
0&0&0&0&0&0&-\sin \theta&\cos \theta\cr }\right ]\Eqno$$

\noi
Within the associative portion $\{e_1,e_2 = e_3 e_4,e_3,e_4\}$ the rotation
indeed remains the same as in the previous example \chietwo, but the
orientation of the rotation in the other two planes is reversed.

Using these ideas, we find that
an appropriate composition of $\chi_{(2,\theta)}, \chi_{(3,4,\theta)}^{(2)},
\chi_{(5,6,\theta)}^{(2)}$, and $\chi_{(7,8,\theta)}^{(2)}$ allows us to
rotate any single plane of the four coordinate planes rotated by
$\chi_{(e_2,\theta)}$.  Notice that $e_3 e_4=e_5 e_6=e_7 e_8=e_2$, i.e.\ the
combinations which appear are all the independent pairs which, in the
multiplication triangle, multiply to the corner $e_2$.  For example,
$\chi_{(2,\theta)} \circ
\chi_{(3,4,\theta)}^{(2)} \circ
\chi_{(5,6,\theta)}^{(2)} \circ \chi_{(7,8,\theta)}^{(2)}$
rotates the
$1$-$2$ plane
by an angle of $4 \theta$.  Similarly,
$\chi_{(2,\theta)} \circ \chi_{(3,4,\theta)}^{(2)} \circ
\chi_{(5,6,-\theta)}^{(2)} \circ \chi_{(7,8,-\theta)}^{(2)}$ rotates the
$3$-$4$ plane by the same amount.

In terms of the multiplication triangle we can give the following rules to
determine the composition needed to do an elementary rotation in the $i$-$j$
plane. Suppose $i=1$, then we need to choose the corner $j$ for the single
$\chi$ and the pairs on the lines leading to $j$ for the three
$\chi^{(2)}$'s.  If neither $i$ nor $j$ is 1, the corner, i.e.\ the single
$\chi$ part, is given by $e_k=e_i e_j$.  The three $\chi^{(2)}$ pieces
come from the pairs which multiply to $e_k$.  The $ij$ piece occurs in the
standard orientation and the other two pairs reversed.

The infinitesimal versions of the two examples above show this structure even
more clearly.  For the first example, $x \mapsto x + \theta\,(e_2x + e_3(e_4x)
+ e_5(e_6x) + e_7(e_8x)) + \Order (\theta^2)$; while for the second example,
$x \mapsto x + \theta\,(e_2x + e_3(e_4x) - e_5(e_6x) - e_7(e_8x)) +
\Order(\theta^2)$.  The infinitesimal version also provides a convenient way
to count the dimension of the group.  There are 7 units and 21 pairs of units
yielding 28 independent generators of $SO(8)$.  As advertised, we have produced
all of $SO(8)$.

As with symmetric multiplication, the Moufang identities \Moufang\ imply that
for any $q,p \in \K_n$, with $|q|=|p|=1$,

$$\chi_q \circ \chi_p \circ \chi_q = \chi_{qpq}\Eqno$$\label\chiqpq

\noi
Therefore we can write any $\chi_{(\rhat,\theta)}$ as a series of flips with
constant angle ${\pi\over 4}$ using \pifourth\ and \chiqpq:

$$\eqalign{
\chi_{(\rhat,\theta)}(x) &:= \exp(\theta\, \rhat)\; x\cr
&\phantom{:}=\exp\left(-{\pi \over 4}\,\shat\right) \left[ \exp\left({\pi
\over 2}(\cos\theta\;\shat + \sin \theta\;\rhat)\right) \left[ \exp\left(-{\pi
\over 4}\,\shat\right) \; x \right] \right]\cr
}\Eqno$$\label\chione

\noi
where $\shat$ is any imaginary unit which anticommutes with $\rhat$.

{}From the second form of $\chi$ we see that  $\Chi$, completely analogously
to $\Psi$  for $\K_8=\O$, maps the same $S^6$ ($\isom\{q \in \O : \, q =
\exp\left({\pi \over 4}\,\rhat\right),\,\Re\rhat = 0,\,|\rhat| =1\}$), now to a
different generating set of $SO(8)$.

Right multiplication is completely analogous to left multiplication.  The
details can easily be worked out using
$xq = \overline{\overline{q}\>\overline{x}}$.

\Section{Lorentz Transformations}

In $(3,1)$ spacetime dimensions, it is standard to use the isomorphism between
$SO(3,1)$ \break and $SL(2,\C)$ to write a vector as a \2by2 hermitian
complex-valued matrix via

$$X^{\mu} \rightarrow X=\pmatrix{x^+ &x^{\phantom{-}}\cr
{\overline{x}}^{\phantom{+}}&x^-\cr}\Eqno$$\label\Lorentz

\noi
where $x^{\pm}=x^0 \pm x^{n+1}\in \R$ are lightcone coordinates,
$x=\sum_{i=1}^n x^i e_i\in \K_n$, and $n=2$.  The Lorentzian norm of $X^{\mu}$
is then given by\Footnote{We use signature $(-1, +1,\dots,+1)$}

$$X^{\mu} X_{\mu} = - {\rm det} X\Eqno$$

\noi
Standard results on determinants of matrices with complex coefficients
show that if $X'$ is obtained from $X$ by the unitary transformation

$$X'= MXM\dag\Eqno$$\label\Trans

\noi
then

$$\eqalign{\det X' = \det (MXM\dag)
&= \det M \det X \det M\dag\cr
&= \det M \det M\dag \det X\cr
&= \vert \det M \vert^2 \det X\cr
&= \det (M M\dag)\, \det X\cr}\Eqno$$\label\Undo

\noi
Therefore, if the determinant of $M$ has norm equal to $1$, then $\det X' =
\det X$ and \Trans\ is a Lorentz transformation.  Notice, however, that
there is some redundancy.  $M$ can be multiplied by an arbitrary overall phase
factor without altering the Lorentz transformation since the phase in $M\dag$
will cancel the phase in $M$.  To remove this redundancy, $M$ is usually
chosen to have determinant equal to $1$ rather than norm $1$, but this
restriction is not necessary.  In Appendix B we record explicit versions of $M$
which give the elementary boosts and rotations.  Any Lorentz transformation
can be obtained from this generating set by doing more than one such
transformation and since

$$X'=(M_n(...(M_1 X M_1\dag)...)M_n\dag)
=(M_n...M_1) X (M_1\dag...M_n\dag)\Eqno$$\label\NewTrans

\noi
we see that {\bf any} finite Lorentz transformation can be implemented by a
single transformation of type \Trans.

We can use \Lorentz, just as in the complex case, to write a vector in
$(n+1,1)$
spacetime dimensions for $n=4,8$ as a \2by2 hermitian matrix with entries in
$K_n$.  The extra quaternionic or octonionic components on the off diagonal
correspond to the extra transverse spatial coordinates.  The manipulations in
\Undo\ are no longer valid in these cases due to the
non-commutativity and non-associativity of the higher dimensional division
algebras, but the last expression on the right hand side is nevertheless equal
to the left hand side.  (Notice that it is also the only expression on the
right hand side which is well-defined.)  A quaternion or octonion valued
matrix $M$ which generates a finite Lorentz transformation in $(n+1,1)$
dimensions must satisfy $\det(M M\dag) =1$.  An octonion valued matrix $M$
must also satisfy an additional restriction which ensures that the
transformation on the right hand side of \Trans\ is well-defined\Footnote{The
conditon that $X'$ be hermitian is identical to the condition that there be no
associativity ambiguity in \Trans.  Both of these things will be true if and
only if $\Im M$ contains only one octonionic direction or if the columns of
$\Im M$ are real multiples of each other.}.

Looking at the elementary boosts and rotations in Appendix B, we see that for
the quaternionic or octonionic cases if we simply let $e_2\rightarrow e_i$, for
$i=2,\dots, n$, then we get all of the new boosts and some of the new
rotations.  The rotations which are missing are just the ones which rotate the
purely imaginary parts of $x$ into each other.  But now consider a
transformation with $M= q {\bf 1} =\exp(\theta\, \rhat) {\bf 1}$, where
$\vert q \vert =1$.  Since the diagonal elements $x^{\pm}$ of $X$ are real,
they are unaffected by these phase transformations.  The off-diagonal
elements, however, transform by a conjugation map:

$$x\mapsto q x \overline{q}\Eqno$$

\noi
As we saw in Section 3, these conjugation maps give all of $SO(3)$ in the
quaternionic case, and if repeated maps are included they give all of $SO(7)$
in the octonionic case.  This is just what we needed.  In the $(3,1)$
dimensional complex case the phase freedom is just the residue left over from
these extra rotations which occur when there is more than one imaginary
direction.

So we have shown that
{\bf all} finite Lorentz transformations can be implemented explicitly as
in \Trans, simply by doing several such transformations in a row:

$$X'=(M_n(...(M_1 X M_1\dag)...)M_n\dag)\Eqno$$\label\NewTrans

Since the octonions are not associative, \NewTrans\ is {\bf not} the same as

$$X'=(M_n...M_1) X (M_1\dag...M_n\dag)\Eqno$$

and it is precisely this non-associativity which means that there is
enough freedom in \NewTrans\ to obtain {\bf any} finite Lorentz transformation.

\Section{Discussion}

First we described $SO(3)$ using quaternions and $SO(7)$ using octonions via
(a series of) conjugation maps, namely the maps $\phi_q$ with
$q=\exp(\theta\,\rhat)$.  We obtained $Aut(O)$ ($\isom G_2$) by restricting
$\theta$ to be ${\pi\over 3}$.  Then we described $SO(4)$ using quaternions
and $SO(8)$ using octonions via the symmetric maps $\psi_q$ and also $SO(8)$
using octonions via left multiplication $\chi_q$.  We suspect that the
existence of two different descriptions of $SO(8)$ is related to triality of
the octonions.

It is worth reiterating here that our implementation of the symmetry groups of
$\H$ and $\O$ provides an interesting new twist on the interpretation of
rotations.  The usual way of looking at a finite rotation is that a fixed axis
is chosen and then the angle of rotation is changed continuously from zero
until the desired rotation is achieved.  Instead, the parameterizations in
terms of flips presented in this paper use building blocks made of rotations
with one fixed angle (${\pi\over 2}$ for $SO(n-1)$ and ${\pi\over 4}$ for
$SO(n)$).  A finite rotation is accomplished by composing several such
rotations, all with the same fixed angle.  The relationship of the various
axes in the composition is varied from initial alignment until the desired
rotation is achieved.  We used these flips to exhibit generating sets for
$SO(8)$, $SO(7)$, and $G_2$ where each generating set is homeomorphic to a
different $S^6$ subset of the octonionic unit sphere $S^7$.  We believe that
the parameterizations in terms of flips are new.  In keeping with this point
of view, the automorphisms of the octonions require flips with constant angle
which is a multiple of ${\pi\over 3}$.

We then used the results for $SO(3)$ and $SO(7)$
to obtain an explicit description of finite Lorentz transformations on
vectors in $(5,1)$ and $(9,1)$ dimensions in terms of unitary transformations
on the $2\times 2$ quaternionic or octonionic matrix representing the vectors.
We believe that the finite version of
$SL(2,\O)$ requiring a succession of such unitary transformations is also new.

A number of other authors have attempted to find similar representations for
the groups we have considered here.  Conway \Ref{J.H.\ Conway, private
communication} has independently developed the finite transformation rules for
$SO(8)$ and $SO(7)$ (without flips), and for $G_2$.  Ramond \Ref{P.\ Ramond,
{\it Introduction to Exceptional Lie Groups and Algebras}, Caltech preprint
CALT-68-577 (1976).}, gives a simple algebraic representation for the finite
elements of $G_2$, $SO(7)$, and $SO(8)$, but uses a mixture of the various
types of multiplication which we have used separately.  A messy representation
for the finite elements of $G_2$ and the infinitesimal elements of $SO(7)$ is
given by G\"unaydin and G\"ursey \Ref{M.\ G\"unaydin and F.\ G\"ursey, {\it
Quark Structure and Octonions,} J.~Math.~Phys.~{\bf 14}. (1973) 1651.}.
Finite transformations were used by Cartan and Schouten \Ref{E. Cartan and
J.A. Schouten, {\it On Riemannian geometries admitting an absolute
parallelism}, Koninklijke Akademie van Wetenschappen te Amsterdam, Proceedings
of the Section of Sciences, {\bf 29}, (1926) 933-946.} to investigate absolute
parallelisms on $S^7$.  Coxeter \Ref{H.S.M.~Coxeter, {\it Integral Cayley
Numbers}, Duke Math.~J.~{\bf 13}, (1946) 561.} gives a special form for
reflections with respect to a hyperplane in $\R^8$.  Infinitesimal
transformations are found more frequently
\Ref{See for example: A. Gamba, {\it Peculiarities of the Eight-Dimensional
Space}, J.~Math. Phys. {\bf 8}, (1967) 4.}.  A detailed analysis can be found
in \Ref{A.R.~D\"underer and F.~G\"ursey, {\it Octonionic representations of
$SO(8)$ and its subgroups and cosets}, J.~Math.~Phys.~{\bf 32}, (1991)
1176.\hfill\break A.R.~D\"underer and F.~G\"ursey, {\it Generalized vector
products, duality and octonionic identities in $D=8$ geometry},
J.~Math.~Phys.~{\bf 25}, (1984) 1496.} where generators of $SO(8)$, $SO(7)$,
and $G_2$ are given in terms of octonions. Their relation to integrated
transformations is indicated but the actual integration is not carried out.

\bigskip
\noi
{\bf Acknowledgements}

\noi
We would like to thank John Conway for his remarks.
CAM would also like to thank Tevian Dray for his suggestions on the finite
Lorentz transformations.  This work was partially funded by NSF grants PHY
89-11757, PHY 92-08494, and DMS 85-05550 (to MSRI).

\bigskip
%\baselineskip=\normalbaselineskip
\References
%\bigskip
%\Figures

\vfill\eject
\baselineskip=\normalbaselineskip

\centerline{\bf APPENDIX A}

\bigskip\bigskip
\noi
Structure matrices for our choice of multiplication rules for the octonions.
(Note that if the sign of the first column is changed, the first matrix
becomes $-{\bf 1}$ and each matrix except the first becomes antisymmetric.)

$$[{\Lambda^1}_{jk}]=\left [\matrix{ 1& 0& 0& 0& 0& 0& 0& 0\cr 0&\phmsp -1& 0&
0& 0& 0& 0& 0\cr 0& 0 &\phmsp -1& 0& 0& 0& 0& 0\cr 0& 0& 0&\phmsp -1& 0& 0& 0&
0\cr 0& 0& 0& 0&\phmsp -1& 0& 0& 0\cr 0& 0& 0& 0& 0&\phmsp -1& 0& 0\cr 0 & 0&
0& 0& 0& 0&\phmsp -1& 0\cr 0& 0& 0& 0& 0& 0& 0&\phmsp -1\cr}\right ]
\quad%$$ $$%
\relax
[{\Lambda^2}_{jk}]=\left [\matrix{ 0& 1& 0& 0& 0& 0& 0& 0\cr 1& 0& 0& 0& 0& 0&
0& 0\cr 0& 0& 0 & 1& 0& 0& 0& 0\cr 0& 0&\phmsp -1& 0& 0& 0& 0& 0\cr 0& 0& 0&
0& 0& 1& 0& 0\cr 0& 0& 0& 0&\phmsp -1& 0& 0& 0\cr 0& 0& 0 & 0& 0& 0& 0& 1\cr
0& 0& 0& 0& 0& 0&\phmsp -1& 0\cr}\right ]$$\medskip $$%
\relax
[{\Lambda^3}_{jk}]=\left [\matrix{ 0& 0& 1& 0& 0& 0& 0& 0\cr 0& 0& 0&\phmsp
-1& 0& 0& 0& 0\cr 1& 0 & 0& 0& 0& 0& 0& 0\cr 0& 1& 0& 0& 0& 0& 0& 0\cr 0& 0&
0& 0& 0& 0&\phmsp -1& 0\cr 0& 0& 0& 0& 0& 0& 0& 1\cr 0& 0& 0 & 0& 1& 0& 0&
0\cr 0& 0& 0& 0& 0&\phmsp -1& 0& 0\cr}\right ]\quad%$$ $$
\relax
[{\Lambda^4}_{jk}]=\left [\matrix{ 0& 0& 0& 1& 0& 0& 0& 0\cr 0& 0& 1& 0& 0& 0&
0& 0\cr 0&\phmsp -1 & 0& 0& 0& 0& 0& 0\cr 1& 0& 0& 0& 0& 0& 0& 0\cr 0& 0& 0&
0& 0& 0& 0& 1\cr 0& 0& 0& 0& 0& 0& 1& 0\cr 0& 0& 0 & 0& 0&\phmsp -1& 0& 0\cr
0& 0& 0& 0&\phmsp -1& 0& 0& 0\cr}\right ]$$\medskip $$
\relax
[{\Lambda^5}_{jk}]=\left [\matrix{ 0& 0& 0& 0& 1& 0& 0& 0\cr 0& 0& 0& 0&
0&\phmsp -1& 0& 0\cr 0& 0 & 0& 0& 0& 0& 1& 0\cr 0& 0& 0& 0& 0& 0& 0&\phmsp
-1\cr 1& 0& 0& 0& 0& 0& 0& 0\cr 0& 1& 0& 0& 0& 0& 0& 0\cr 0& 0&
\phmsp -1& 0& 0& 0& 0& 0\cr  0& 0& 0& 1& 0& 0& 0& 0\cr}\right ]\quad%$$ $$
\relax
[{\Lambda^6}_{jk}]=\left [\matrix{ 0& 0& 0& 0& 0& 1& 0& 0\cr 0& 0& 0& 0& 1& 0&
0& 0\cr 0& 0& 0 & 0& 0& 0& 0&\phmsp -1\cr 0& 0& 0& 0& 0& 0&\phmsp -1& 0\cr
0&\phmsp -1& 0& 0& 0& 0& 0& 0\cr 1& 0& 0& 0& 0& 0& 0& 0\cr 0& 0& 0 & 1& 0& 0&
0& 0\cr 0& 0& 1& 0& 0& 0& 0& 0\cr}\right ]$$\medskip $$
\relax
[{\Lambda^7}_{jk}]=\left [\matrix{ 0& 0& 0& 0& 0& 0& 1& 0\cr 0& 0& 0& 0& 0& 0&
0&\phmsp -1\cr 0& 0 & 0& 0&\phmsp -1& 0& 0& 0\cr 0& 0& 0& 0& 0& 1& 0& 0\cr 0&
0& 1& 0& 0& 0& 0& 0\cr 0& 0& 0&\phmsp -1& 0& 0& 0& 0\cr 1& 0 & 0& 0& 0& 0& 0&
0\cr 0& 1& 0& 0& 0& 0& 0& 0\cr}\right ]\quad%$$ $$
\relax
[{\Lambda^8}_{jk}]=\left [\matrix{ 0& 0& 0& 0& 0& 0& 0& 1\cr 0& 0& 0& 0& 0& 0&
1& 0\cr 0& 0& 0 & 0& 0& 1& 0& 0\cr 0& 0& 0& 0& 1& 0& 0& 0\cr 0& 0& 0&\phmsp
-1& 0& 0& 0& 0\cr 0& 0&\phmsp -1& 0& 0& 0& 0& 0\cr 0&\phmsp -1& 0 & 0& 0& 0&
0& 0\cr 1& 0& 0& 0& 0& 0& 0& 0\cr}\right ] $$

\vfill\eject
\centerline{\bf APPENDIX B}

\bigskip\bigskip
\noi Using the following correspondence, which is explained in Section 5:
$$X^{\mu} \longleftrightarrow X=\pmatrix{x^+&x^{\phantom{-}}\cr
{\overline{x}}^{\phantom{+}}&x^-\cr}$$
\noi we can write the elementary Lorentz transformations ${L^{\mu}}_{\nu}$ in
terms of \2by2 hermitian matrices $M$ over $\K_n$.
$$X'^{\mu}={L^{\mu}}_{\nu} X^{\nu} \longleftrightarrow
X'=\left\lbrace \matrix{M X M\dag,\hfill&{\rm for\ {\it Categories\ 1}\ and\
{\it 2}}\hfill\cr
\noalign{\medskip}
M_2\left(M_1 X M_1\dag \right)M_2\dag\,,\hfill&{\rm for\ {\it Category\
3}}\hfill\cr}\right.$$

\bigskip
\noi{\it Category 1}: Boosts
$$\eqalignno{
\noalign{\hbox{$X^0 \leftrightarrow X^1$:}\medskip}
L=\pmatrix{\cosh\alpha&\sinh\alpha&0&\ldots&0\cr
\sinh\alpha&\cosh\alpha&0&\ldots&0\cr
0&0&1&\ldots&0\cr \vdots&\vdots&\vdots&\ddots&\vdots\cr 0&0&0&\ldots&1\cr}
&\longleftrightarrow
M=\pmatrix{\cosh\left({\alpha \over 2}\right)&\sinh\left({\alpha \over
2}\right)\cr\noalign{\smallskip}\sinh\left({\alpha \over
2}\right)&\cosh\left({\alpha \over
2}\right)\cr}\cr
\noalign{\bigskip\bigskip}
\noalign{\hbox{$X^0 \leftrightarrow X^i$:}\medskip}
L=\pmatrix{\cosh\alpha&0&\ldots&0&\sinh\alpha&0&\ldots&0\cr
0&1&\ldots&0&0&0&\ldots&0\cr
\vdots&\vdots&\ddots&\vdots&\vdots&\vdots&\ddots&\vdots\cr
0&0&\ldots&1&0&0&\ldots&0\cr
\sinh\alpha&0&\ldots&0&\cosh\alpha&0&\ldots&0\cr
0&0&\ldots&0&0&1&\ldots&0\cr
\vdots&\vdots&\ddots&\vdots&\vdots&\vdots&\ddots&\vdots\cr
0&0&\ldots&0&0&0&\ldots&1\cr}&\longleftrightarrow
M=\pmatrix{\phantom{-e_i\,} \cosh\left({\alpha \over
2}\right)&e_i\,\sinh\left({\alpha \over 2}\right)\cr\noalign{\smallskip}
-e_i\,\sinh\left({\alpha \over 2}\right)&\phantom{e_i\,} \cosh\left({\alpha
\over 2}\right)\cr}\cr
\noalign{\bigskip\bigskip}
\noalign{\hbox{$X^0 \leftrightarrow X^n$:}\medskip}
L=\pmatrix{\cosh\alpha&0&\ldots&0&\sinh\alpha\cr
0&1&\ldots&0&0\cr \vdots&\vdots&\ddots&\vdots\cr 0&0&\ldots&1&0\cr
\sinh\alpha&0&\ldots&0&\cosh\alpha\cr}&\longleftrightarrow
M=\pmatrix{\exp\left({\alpha \over 2}\right)&0\cr\noalign{\smallskip}
0&\exp\left(-{\alpha \over 2}\right)\cr}\cr}$$
\vfill\eject

\vsize=9.1in\null\vskip-.5in
\noi {\it Category 2}: Rotations
$$\eqalignno{
\noalign{\hbox{$X^1 \leftrightarrow X^i$:}\smallskip}
L=\pmatrix{1&0&0&\ldots&0&0&0&\ldots&0\cr
0&\cos\alpha&0&\ldots&0&-\sin\alpha&0&\ldots&0\cr
0&0&1&\ldots&0&0&0&\ldots&0\cr
\vdots&\vdots&\vdots&\ddots&\vdots&\vdots&\vdots&\ddots&\vdots\cr
0&0&0&\ldots&1&0&0&\ldots&0\cr
0&\sin\alpha&0&\ldots&0&\phantom{-}\cos\alpha&0&\ldots&0\cr
0&0&0&\ldots&0&0&1&\ldots&0\cr
\vdots&\vdots&\vdots&\ddots&\vdots&\vdots&\vdots&\ddots&\vdots\cr
0&0&0&\ldots&0&0&0&\ldots&1\cr}&\longleftrightarrow
M=\pmatrix{\exp\left(e_i{\alpha \over 2}\right)&0\cr\noalign{\smallskip}
0&\exp\left(-e_i{\alpha \over 2}\right)\cr}\cr
\noalign{\medskip}
\noalign{\hbox{$X^1 \leftrightarrow X^n$:}\smallskip}
L=\pmatrix{1&0&0&\ldots&0&0\cr
0&\cos\alpha&0&\ldots&0&-\sin\alpha\cr
0&0&1&\ldots&0&0\cr
\vdots&\vdots&\vdots&\ddots&\vdots&\vdots\cr
0&0&0&\ldots&1&0\cr
0&\sin\alpha&0&\ldots&0&\phantom{-}\cos\alpha\cr}&\longleftrightarrow
M=\pmatrix{\phantom{-}\cos\left({\alpha \over 2}\right)&\sin\left({\alpha \over
2}\right)\cr\noalign{\smallskip} -\sin\left({\alpha \over
2}\right)&\cos\left({\alpha \over
2}\right)\cr}\cr
\noalign{\medskip}
\noalign{\hbox{$X^i \leftrightarrow X^n$:}\smallskip}
L=\pmatrix{1&\ldots&0&0&0&\ldots&0&0\cr
\vdots&\ddots&\vdots&\vdots&\vdots&\ddots&\vdots&\vdots\cr
0&\ldots&1&0&0&\ldots&0&0\cr
0&\ldots&0&\cos\alpha&0&\ldots&0&-\sin\alpha\cr
0&\ldots&0&0&1&\ldots&0&0\cr
\vdots&\ddots&\vdots&\vdots&\vdots&\ddots&\vdots&\vdots\cr
0&\ldots&0&0&0&\ldots&1&0\cr
0&\ldots&0&\sin\alpha&0&\ldots&0&\phantom{-}\cos\alpha\cr}&\longleftrightarrow
M=\pmatrix{\phantom{e_i\,} \cos\left({\alpha \over
2}\right)&e_i\,\sin\left({\alpha \over 2}\right)\cr\noalign{\smallskip}
e_i\,\sin\left({\alpha \over 2}\right)&\phantom{e_i\,}
\cos\left({\alpha \over 2}\right)\cr}\cr}$$

\medskip
\noi {\it Category 3}: Additional Transverse Rotations
\medskip
\noi $X^i \leftrightarrow X^j$:\vskip -\bigskipamount
{\Small
$$\eqalignno{
L=\pmatrix{1&\ldots&0&0&0&\ldots&0&0&0&\ldots&0\cr
\vdots&\ddots&\vdots&\vdots&\vdots&\ddots&\vdots&\vdots&\vdots&\ddots&\vdots\cr
0&\ldots&1&0&0&\ldots&0&0&0&\ldots&0\cr
0&\ldots&0&\cos\alpha&0&\ldots&0&-\sin\alpha&0&\ldots&0\cr
0&\ldots&0&0&1&\ldots&0&0&0&\ldots&0\cr
\vdots&\ddots&\vdots&\vdots&\vdots&\ddots&\vdots&\vdots&\vdots&\ddots&\vdots\cr
0&\ldots&0&0&0&\ldots&1&0&0&\ldots&0\cr
0&\ldots&0&\sin\alpha&0&\ldots&0&\phantom{-}\cos\alpha&0&\ldots&0\cr
0&\ldots&0&0&0&\ldots&0&0&1&\ldots&0\cr
\vdots&\ddots&\vdots&\vdots&\vdots&\ddots&\vdots&\vdots&\vdots&\ddots&\vdots\cr
0&\ldots&0&0&0&\ldots&0&0&0&\ldots&1\cr
}&\longleftrightarrow
\eqalign{
M_1&= \exp\left(-{\pi \over 2} e_i\right)
\pmatrix{1&0\cr\noalign{\smallskip} 0&1\cr}\cr
M_2& = \exp\left({\pi \over 2}\left(\cos{\alpha\over 2}\; e_i +
\sin{\alpha\over 2}\;
e_j\right)\right)\pmatrix{1&0\cr\noalign{\smallskip} 0&1\cr}\cr}\cr
}$$}

\bye